\documentclass[conference]{IEEEtran}
\usepackage{algorithm,algorithmic}
\makeatletter
\newcommand\fs@betterruled{%
  \def\@fs@cfont{\bfseries}\let\@fs@capt\floatc@ruled
  \def\@fs@pre{\vspace*{5pt}\hrule height.8pt depth0pt \kern2pt}%
  \def\@fs@post{\kern2pt\hrule\relax}%
  \def\@fs@mid{\kern2pt\hrule\kern2pt}%
  \let\@fs@iftopcapt\iftrue}
\floatstyle{betterruled}
\restylefloat{algorithm}
\makeatother
\newcommand{\squeezeup}{\vspace{-4mm}}
\newcommand{\squeezeupan}{\vspace{-2mm}}
\newcommand{\squeezeupann}{\vspace{-1mm}}

\usepackage{cite}
\usepackage{amsmath,amssymb,amsfonts}
\usepackage{algorithmic}
\usepackage{graphicx}
\usepackage{textcomp}
\usepackage{comment}
\usepackage{enumitem}
\usepackage{xcolor}
\def\BibTeX{{\rm B\kern-.05em{\sc i\kern-.025em b}\kern-.08em
    T\kern-.1667em\lower.7ex\hbox{E}\kern-.125emX}}
\renewcommand{\a}{\mathbf{a}}

\renewcommand{\c}{\mathbf{c}}
\renewcommand{\d}{\mathbf{d}}

\newcommand{\h}{\mathbf{h}}

\newcommand{\n}{\mathbf{n}}

\renewcommand{\t}{\mathbf{t}}
\renewcommand{\u}{\mathbf{u}}
\renewcommand{\v}{\mathbf{v}}

\newcommand{\y}{\mathbf{y}}

\renewcommand{\H}{\mathbf{H}}

\newcommand{\R}{\mathbf{R}}

\newcommand{\U}{\mathbf{U}}

\newcommand{\Y}{\mathbf{Y}}

\begin{document}
\title{Direction-Assisted Beam Management in Full Duplex Millimeter Wave Massive MIMO Systems}

\author{\IEEEauthorblockN{Md Atiqul Islam\IEEEauthorrefmark{2}, George C. Alexandropoulos\IEEEauthorrefmark{4}, and Besma Smida\IEEEauthorrefmark{2}}
\IEEEauthorblockA{{\IEEEauthorrefmark{2}Department of Electrical and Computer Engineering, University of Illinois at Chicago, USA}\\
\IEEEauthorrefmark{4}Department of Informatics and Telecommunications, National and Kapodistrian University of Athens, Greece\\
emails: \{mislam23,smida\}@uic.edu, alexandg@di.uoa.gr
}}
\maketitle

\begin{abstract}
Recent applications of the Full Duplex (FD) technology focus on enabling simultaneous control communication and data transmission to reduce the control information exchange overhead, which impacts end-to-end latency and spectral efficiency. In this paper, we present a simultaneous direction estimation and data transmission scheme for millimeter Wave (mmWave) massive Multiple-Input Multiple-Output (MIMO) systems, enabled by a recent FD MIMO technology with reduced hardware complexity Self-Interference (SI) cancellation. We apply the proposed framework in the mmWave analog beam management problem, considering a base station equipped with a large transmit antenna array realizing downlink analog beamforming and few digitally controlled receive antenna elements used for uplink Direction-of-Arrival (DoA) estimation. A joint optimization framework for designing the DoA-assisted analog beamformer and the analog as well as digital SI cancellation is presented with the objective to maximize the achievable downlink rate. Our simulation results showcase that the proposed scheme outperforms its conventional half-duplex counterpart, yielding reduced DoA estimation error and superior downlink data rate.
\end{abstract}


\section{Introduction}
In-band Full Duplex (FD) massive Multiple-Input Multiple-Output (MIMO) communication is a promising technology for the increasingly demanding data rate requirements of future wireless networks due to its inherent capability to enable simultaneous UpLink (UL) and DownLink (DL) communications within the entire frequency band \cite{Samsung,B:Full-Duplex,alexandropoulos2017joint}. In addition to concurrent UL/DL data transmission in reciprocal communications, the FD technology has been recently considered for simultaneous DL data transmission and UL Channel State Information (CSI) estimation, where FD Base Stations (BSs) transmit data symbols in the DL via CSI-dependent multi-user BeamForming (BF), while receiving UL training symbols from the user terminals \cite{islam2020simultaneous,mirza2018performance,Islam_2020_Sim_Multi}. The principal bottleneck of any FD system is the in-band Self-Interference (SI) signal at the Receiver (RX) side, resulting from the simultaneous transmission and reception. Recently in \cite{xiao2017full_all,Vishwanath_2020,alexandropoulos2020full}, FD MIMO systems operating at millimeter Wave (mmWave) frequencies were introduced, where SI suppression was achieved through a combination of propagation domain isolation, analog domain suppression, and digital SI cancellation techniques.

Wireless communications at mmWave frequencies are mainly realized via highly directional BF, which is enabled by massive MIMO transceivers \cite{bjornson2016massive} that are capable of analog or hybrid Analog and Digital (A/D) beamforming \cite{vlachos2019}. To alleviate the hardware cost in those transceiver architectures, the large-scale antenna arrays are connected to small numbers of Radio Frequency (RF) chains via analog preprocessing networks comprised of phase shifters \cite{venkateswaran2010analog}. However, the selection of the adequate analog beams from the predefined codebooks and the design of the digital beamformers require CSI knowledge, which is hard to acquire with realistically affordable latency, especially under mobility conditions (see, e.g.,\cite{ULBA2021} and references therein). This challenge will be naturally more pronounced in the envisioned FD massive MIMO systems operating at mmWave frequencies, where multiple UL and DL channels need to be estimated. In practice, the analog BF in mmWave massive MIMO systems is designed via beam switching between the communicating nodes in order to find a pair of beams from their available beam codebooks meeting a link performance indicator threshold \cite{BMtutorial,aykin2020efficient}. However, such time-consuming beam training procedures incur significant configuration overhead deprived of data transmission. Hence, beam misalignment may yield poor performance under mobility scenarios \cite{palacios2017tracking}. Alternatively, exploiting UL/DL reciprocity or position information, the DL BF can be performed in the direction of the UL dominant Direction-of-Arrival (DoA), thus, reducing the beam sweeping overhead \cite{bjornson2016massive,alexandropoulos2017position}. 
\begin{figure*}[!tpb]
	\begin{center}
	\includegraphics[width=0.97\linewidth]{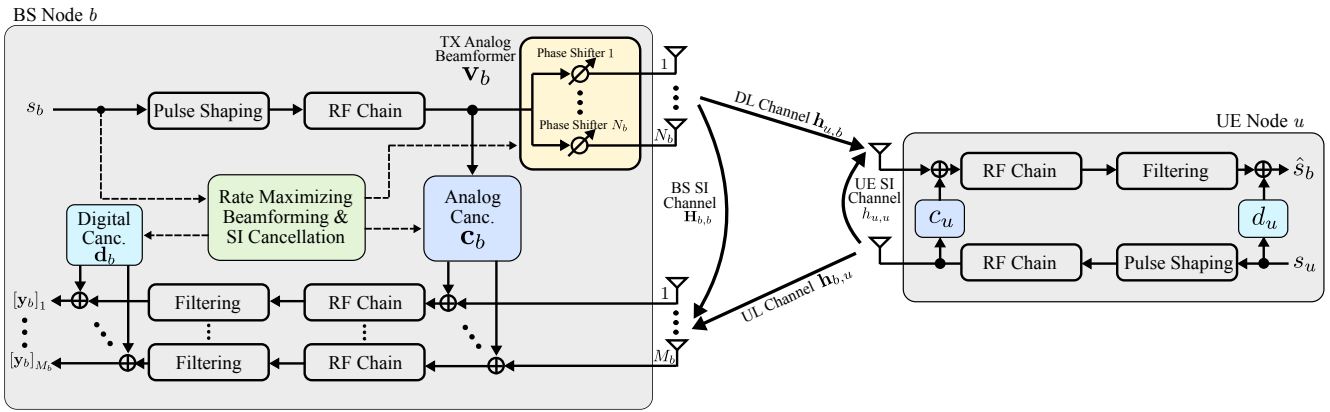}
	\caption{The considered FD massive MIMO communication system operating at mmWave frequencies: an FD massive MIMO Base Station (BS) capable of analog BF communicates with a mobile single-antenna FD User (UE), where the latter node transmits training signals in the UL for DoA estimation.}
	\label{fig: FD_HBF_Sys}
	\end{center}
	\squeezeup
	\squeezeupann
\end{figure*}

In this paper, we present a DoA-assisted beam management framework for FD mmWave massive MIMO systems, where the BS is equipped with a large antenna array realizing DL analog BF and few digitally controlled receive antenna elements used for UL DoA estimation. Capitalizing on the recently presented FD hardware architecture for hybrid A/D BF in \cite{alexandropoulos2020full}, we propose a Simultaneous DoA estimation and Data Transmission (SDDT) scheme for boosting beam management in FD mmWave massive MIMO communications. Enabled by FD and leveraging channel reciprocity, we simultaneously estimate the UL dominant DoA and transmit analog beamformed data in the DL direction. We present a joint design of the DoA-assisted analog BF as well as the A/D SI cancellation units, targeting the maximization of the achievable DL rate. Our extensive simulation results considering a mmWave channel model showcase the FD-enabled gains of DoA-assisted beam management under various user mobility conditions. In particular, the proposed FD-based SDDT scheme achieves approximately $1.2\times$ the DL rate of conventional Half Duplex (HD) mmWave massive MIMO systems. 

\section{System and Signal Models}
We consider an FD mmWave massive MIMO Base Station (BS) node $b$ equipped with $N_b$ TX and $M_b$ RX antenna elements communicating with a mobile single-antenna FD User Equipment (UE) node $u$. In particular, in the TX of the BS node $b$, a single RF chain is connected via phase shifters with a large antenna array of $N_b$ elements, whereas each of the few $M_b$ RX antennas is connected to a dedicated RF chain. In this work, a small number for $M_b$ is sufficient for UL DoA estimation; extension for large $M_b$ and analog combining is left for future work. Using the values of the $N_b$ phase shifters, we can formulate the TX analog beamforming vector $\v_b \in \mathbb{C}^{N_b\times 1}$, whose elements are assumed to have constant magnitude, i$.$e$.$, $|[\v_b]_{n}|^2=1/N_b$ $\forall n=1,2,\ldots,N_b$. Besides, we assume that the angles of the analog phase shifters are quantized and have a finite set of possible values. Therefore,
$\v_b\in\mathbb{F}_{\rm TX}$, where $\mathbb{F}_{\rm TX}$ represents the predefined beam codebook including ${\rm card}(\mathbb{F}_{\rm TX})$ distinct vectors (or analog beams) \cite{gonzalez2018channel,ULBA2021}. We assume that, for every channel use, the BS node $b$ transmits the complex-valued information data symbol $s_b$ (chosen from a discrete modulation set) after analog BF with $\v_b$. Similarly, the single-antenna FD UE node $u$ sends the training symbol ${s}_u$ through the UL channel. The signal transmissions at both node $b$ and node $u$ are power limited according to $\mathbb{E}\{\|\mathbf{v}_b{s}_b\|^2\}\leq {\rm P}_b$ and $\mathbb{E}\{|{s}_u|^2\}\leq {\rm P}_u$, respectively.

\subsection{Channel Model}
All the considered mmWave UL/DL channels consist of multiple propagation paths. In the spatial domain, each propagation path can be characterized by its Direction-of-Arrival (DoA)/ Direction-of-Departure (DoD) and the corresponding power as well as phase components \cite{akdeniz2014millimeter}. For a certain coherence block, the UL channel $\h_{b,u} \in \mathbb{C}^{M_b \times 1}$ including a Line-of-Sight (LoS) component and $L_p\!-\!1$ non-Line-of-Sight (nLoS) paths is mathematically expressed as
\begin{equation}\label{eq: UL_chan}
    \begin{split}
        \h_{b,u} \triangleq  \beta_{\rm LoS} \a_{M_b}(\theta_{\rm LoS}) + \sum\limits_{\ell = 1}^{L_p-1} \beta_{{\rm nLoS},\ell} \a_{M_b}(\theta_{{\rm nLoS},\ell}),
    \end{split}
\end{equation}
where $\beta_{\rm LoS}\in \mathbb{C}$ and $\theta_{\rm LoS}\in [0\,\, 2\pi]$ represent the complex gain and the DoA of the LoS path, respectively. Here, $\beta_{{\rm nLoS},\ell}\in \mathbb{C}$ and $\theta_{{\rm nLoS},\ell}\in[0\,\, 2\pi]$~$\forall \ell = \{1,2,\ldots, L_p-1\}$ are the complex gains and the DoAs of the nLoS path, respectively. Considering a Uniform Linear Array (ULA), the response vector $\a_{M_b}(\theta)$ for $M_b$ antenna elements and any DoA $\theta$ is formulated as \cite{gonzalez2018channel}
\begin{equation}
    \begin{split}
        \a_{M_b}(\theta)\! \triangleq \! \frac{1}{\sqrt{M_b}} \big[1,e^{j\frac{2\pi}{\lambda}d \sin(\theta)},\ldots,e^{j\frac{2\pi}{\lambda}(M_b-1)d \sin(\theta)}]^{\rm T},
    \end{split}
\end{equation}
where $\lambda$ is the propagation signal wavelength and $d$ denotes the distance between adjacent antenna elements. 

For the considered FD system, the UL and DL channels are reciprocal resulting in similar complex gains and DoAs/DoDs  \cite{gonzalez2018channel,bjornson2016massive}. Therefore, similar to \eqref{eq: UL_chan}, the DL channel $\h_{u,b}^{\rm T}\in \mathbb{C}^{1 \times N_b}$ can be expressed as follows:
\begin{equation}
    \begin{split}
        \h_{u,b}^{\rm T} \triangleq  \beta_{\rm LoS} \a_{N_b}^{\rm H}(\theta_{\rm LoS}) + \sum\limits_{\ell = 1}^{L_p-1} \beta_{{\rm nLoS},\ell} \a_{N_b}^{\rm H}(\theta_{{\rm nLoS},\ell}).
    \end{split}
\end{equation}
In this paper, our goal is to estimate the LoS DoA $\theta_{\rm LOS}$ using the UL training symbols and find the DoA-assisted analog beamformer $\v_b$ that maximizes the DL rate. However, due to the FD operation at both nodes, the simultaneous DL data and UL training transmission induce SI signal in the RXs of the BS and UE. Following \cite{alexandropoulos2020full,satyanarayana2018hybrid}, we consider the mmWave clustered model for the SI channels denoted by $\mathbf{H}_{b,b} \in \mathbb{C}^{M_b\times N_b}$ for the BS node $b$ and ${h}_{u,u}\in\mathbb{C}$ for the UE node $u$, where $\kappa$ represent the Rician factor.

\squeezeupan
\subsection{Received Signal Model}
At the RX of the BS node $b$, the training symbols transmitted from the UE node $u$ are received along with the SI signal induced by the simultaneous data and training signal transmissions. Due to limited propagation attenuation, the strong SI signal is capable of driving the RX RF chains into saturation. Therefore, as shown in Fig.~\ref{fig: FD_HBF_Sys}, an $N$-tap analog SI canceller is utilized to suppress the SI at the inputs of the RX RF chains, which is based on the TX RF chain output. Exploiting the low-complexity cancellation in \cite{alexandropoulos2020full}, we employ the canceller taps at the TX RF chain output at the BS node $b$ resulting in an analog SI canceller where the number of taps does not scale with the number of antenna elements. To suppress the residual SI signal after analog cancellation, a digital SI canceller is utilized at the BS baseband. Denoting $\mathbf{c}_{b}, \d_{b} \in\mathbb{C}^{M_b\times 1}$  as the analog and digital SI cancellers, respectively, the baseband received signal at the BS node $b$, $\mathbf{y}_{b}\in\mathbb{C}^{M_b\times 1}$, is given by
\begin{equation}\label{eq: sig_yb}
    \begin{split}
        \y_b &\triangleq \h_{b,u} s_u + (\H_{b,b}\v_b + \c_b + \d_b) s_b + \n_b,
    \end{split}
\end{equation}
where $\n_{b}\in \mathbb{C}^{M_b\times 1}$ is the zero-mean Additive White Gaussian Noise (AWGN) with variance $\sigma^2_{b}\mathbf{I}_{M_b}$. 
Similarly, the DL signal $y_u\in\mathbb{C}$ received at the UE node $u$ after analog and digital SI cancellation can be expressed as
\begin{equation}
    \begin{split}
        y_u &\triangleq \h_{u,b}^{\rm T}\v_b s_b + ({h}_{u,u} + c_u + d_u) s_u + n_u,
    \end{split}
\end{equation}
where $n_u\in \mathbb{C}$ represents the AWGN with variance $\sigma^2_{u}$.
As previously mentioned, for proper FD-based reception, the RXs' RF chains need to be unsaturated from any residual SI stemming out analog SI cancellation at both nodes $b$ and $u$ \cite{alexandropoulos2017joint, islam2019unified}. Denoting the residual SI power thresholds as $\lambda_b$ and $\lambda_u$ at node $b$ and node $u$, respectively, the residual SI power constraints ${\rm P}_b|[({\H}_{b,b}\v_b+\c_{b})]_{(j,:)}|^2\!\leq \!\lambda_b \,\forall j=1,2,\ldots,M_b$ and ${\rm P}_{u}|({h}_{u,u} + c_u)|^2\leq \lambda_u$ are necessary to be satisfied for successful reception after analog SI suppression.

\section{DoA Estimation and DL Data Transmission}
In this section, we introduce the proposed DoA-assisted analog beam management protocol along with the DoA tracking scheme for the considered UL/DL mmWave channel with the proposed FD massive MIMO system.
\squeezeupann
\subsection{UL/DL Channel Evolution Properties}
We assume wireless communications in Time Division Duplexing (TDD) manner, where the considered channels remain constant for all the channel uses in a time slot, and the channel properties (i.e., DoAs/DoDs, complex path gains) of the successive time slots are temporally correlated. Each time slot of $T_s$ time units contains $L$ symbols (i.e., $L$ channel uses). For any consecutive $(i\!-1\!)$ and $i$ time slots, the evolution of LoS DoA component is expressed similar to \cite{va2016beam} as
\begin{equation}
    \begin{split}
        \theta_{\rm LoS}[i] \triangleq \theta_{\rm LoS}[i-1] + \Delta\theta,
    \end{split}
\end{equation}
where $\Delta\theta$ depends on the velocity of the UE node $u$ and the time slot duration $T_s$. We also assume that the UE node is moving with the constant velocity $v$, hence, $\Delta\theta \triangleq \text{arctan}\left(\frac{vT_s}{d_{\rm BS}}\right)$ with $d_{\rm BS}$ being the distance between the UE and the BS nodes.
The evolution model for the LoS complex path gain $\beta_{\rm LOS}$ is given by the first-order Gauss-Markov model as \cite{va2016beam} 
\begin{equation}
    \begin{split}
        \beta_{\rm LoS}[i] \triangleq \rho\beta_{\rm LoS}[i-1] + \epsilon[i-1],
    \end{split}
\end{equation}
where $\rho$ is the correlation coefficient, and $\epsilon[i-1]$ is the zero-mean
complex Gaussian noise distributed as $\mathcal{CN}(0,1-\rho^2)$. The DoAs and complex path gains of the nLoS components are assumed to change randomly between consecutive time slots. In this paper, we propose to track only the LoS DoA component for each time slot, and then use it analog beam selection. The estimation of all complex path gains and their correlation coefficients is left for future investigation.

\begin{figure}
    \centering
    \includegraphics[width=0.9\linewidth]{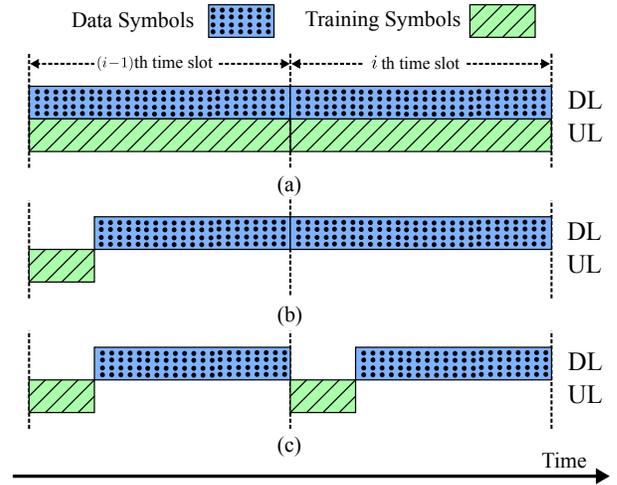}
    \caption{(a) Proposed FD SDDT Scheme, (b) Conventional HD DoA with update after threshold, and (c) Conventional HD DoA estimation in each slot.}
    \label{fig: trans_protocol}
\end{figure}

\subsection{Proposed FD-Based SDDT Protocol}
The proposed protocol for simultaneous DoA estimation and data transmission is illustrated in Fig.~\ref{fig: trans_protocol}(a), where the DL channel is dedicated for data transmission, while the UL is accessed to transmit training symbols. The procedures of DoA-based tracking and transmission are described as follows:
\begin{enumerate}
    \item DL Data transmission at any $i$th time slot leverages the DOA estimation for the previous $(i\!-\!1)$th time slot, which is realized via simultaneous UL transmission of $L$ orthogonal training symbols sent from the UE node $u$.
    \item Capitalizing on the UL/DL reciprocity, any LoS component estimation $\widehat{\theta}_{\rm LoS}[i-1]$ is used during the next $i$th time slot for the DL data transmission precoding. 
    \item The DL analog beamformer $\v_b[i]\in\mathbb{F}_{\rm TX}$ at each $i$th time slot is chosen according to $\widehat{\theta}_{\rm LoS}[i-1]$ subject to the satisfaction of the residual SI power constraints.
\end{enumerate}
In Fig.~\ref{fig: trans_protocol}(b), a conventional HD DoA-assisted analog beamforming scheme is illustrated, where, contrary to the FD case, a fraction of the available symbols in a time slot is dedicated for DoA estimation. Once the DoA of a time slot is estimated, its analog beamforming and that of the succeeding ones utilizes the same DoA estimation, unless the received Signal-to-Noise Ratio (SNR) falls below a certain threshold; in this case, a DoA estimation update occurs. An HD data transmission protocol with DoA estimation at each time slot is shown in Fig.~\ref{fig: trans_protocol}(c).

\subsection{UL DoA Estimation}
As previously discussed, the UL training symbols at any $(i\!-\!1)$th time slot are used for this slot's LoS DoA estimation, which is leveraged for analog BF at the $i$th time slot. For this estimation, we deploy the MUltiple SIgnal Classification (MUSIC) algorithm; other DoA estimation techniques can be used as well \cite{krim1996two}. Starting from \eqref{eq: sig_yb}, the received $L$ training symbols at any $(i\!-\!1)$th time slot can be grouped in $\mathbf{Y}_{b}[i\!-\!1]\in \mathbb{C}^{M_b\times L}$. For the MUSIC DoA estimation, the received signal covariance matrix $\R_{b}\in\mathbb{C}^{M_b\times M_b}$ can be estimated as
\begin{equation}\label{eq: sampl_cov}
    \begin{split}
        {\R}_b\! \triangleq\! \mathbb{E}\{\Y_b[i\!-\!1]\Y_b[i\!-\!1]^{\rm H}\}\! ,\, \widehat{\R}_b\! \triangleq \!\frac{1}{L}  \Y_b[i\!-\!1]\Y_b^{\rm H}[i\!-\!1].
    \end{split}
\end{equation}
By taking the eigenvalue decomposition of the estimated sample covariance matrix $\widehat{\R}_b$, it is deduced that:
\begin{equation}\label{eq: cov_eig}
    \begin{split}
        \widehat{\R}_b \triangleq \U {\rm diag}\{\eta_1,\eta_2,\ldots,\eta_{M_b}\}\U^{\rm H},
    \end{split}
\end{equation}
where $\eta_1\geq\eta_2\geq\ldots\geq\eta_{M_b}$ are the eigenvalues of $\widehat{\R}_b$ and $\U\in\mathbb{C}^{M_b\times M_b}$ contains their corresponding eigenvectors. Since we are interested in estimating the DoA of the LoS component, the matrix $\U$ can be partitioned as $\U=[\u_s|\U_n]$, where the columns in $\U_{n}\in\mathbb{C}^{M_b\times M_b-1}$ are the eigenvectors spanning the noise subspace and $\u_s$ is the signal space eigenvector. We next project the search vector $\a_{M_b}(\theta)$~$\forall \theta\in[0\,\,2\pi]$ onto the noise subspace $\U_{n}$ and calculate the spectral peak as
\begin{equation}\label{eq: spectral_peak}
    \begin{split}
        S(\theta) \triangleq \frac{1}{\a_{M_b}^{\rm H}(\theta)\U_{n}\U_{n}^{\rm H}\a_{M_b}(\theta)}.
    \end{split}
\end{equation}
Finally, the position $\theta$ of the spectral peak is the estimated LoS DoA $\widehat{\theta}_{\rm LoS}[i\!-\!1]$ corresponding to this $(i\!-\!1)$th time slot. 

\subsection{DoA-Assisted Analog Beamforming and DL Rate}
Capitalizing on the UL/DL reciprocity and the estimated LoS DoA component, we propose to approximate the DL channel at each $i$th time slot as follows:
\begin{equation}\label{eq: DL_est_chan}
    \begin{split}
        \widehat{\h}_{u,b}^{\rm T}[i] \triangleq \a_{N_b}^{\rm H}(\widehat{\theta}_{\rm LoS}[i-1]).
    \end{split}
\end{equation}
This approximation is further used during this $i$th time slot to find the best analog beamformer $\v_{b}[i]$ at the BS node $b$, via searching in the available beam codebook $\mathbb{F}_{\rm TX}$. 
Therefore, the instantaneous achievable DL rate per channel use for the proposed FD-based SDDT scheme at this time slot is given by
\begin{equation}
    \begin{split}\label{eq:rate}
        \mathcal{R}_{\rm DL} [i] = \log_2\left(1+ \frac{{\rm P}_{b}|\widehat{\h}_{u,b}^{\rm T}[i]\v_{b}[i]|^2}{\sigma_u^2 + \sigma_{r,u}^2[i]} \right),
    \end{split}
\end{equation}
where $\sigma_{r,u}^2[i] \triangleq {\rm P}_{u}|({h}_{u,u}[i] + c_u[i] + d_u[i])|^2$ is the residual signal power after both A/D SI cancellation at the UE node $u$.

\section{Proposed SDDT Optimization Framework}
In this section, we focus on the joint design of the analog beamformer $\v_b$, the analog SI cancellers $\c_b$ and $c_u$, as well as the digital canceller $\d_b$ and $d_u$ at the BS node $b$ and the UE node $u$, which maximize the estimated achievable DL rate in \eqref{eq:rate} at each $i$th time slot. Based on the availability of the DL channel estimation $\widehat{\h}_{u,b}[i]$, as well as the SI estimations $\widehat{\H}_{b,b}[i]$ and $\widehat{h}_{u,u}[i]$, we consider the optimization problem:
\begin{align}\label{eq: optimization_eq}
        \nonumber\underset{\substack{\v_b[i],\c_b[i], \d_{b}[i]\\ c_u[i], d_u[i]}}{\text{max}} &\log_2\left(1+ \frac{{\rm P}_{b}|\widehat{\h}_{u,b}^{\rm T}[i]\v_{b}[i]|^2}{\sigma_u^2 + \sigma_{r,u}^2[i]} \right)\\
        \text{\text{s}.\text{t}.}\quad
        &\!\!\!{\rm P}_b|[(\widehat{\H}_{b,b}[i]\v_b[i]\!+\!\c_{b}[i])]_{(j,:)}|^2\!\leq \!\lambda_b \!\,\forall j\!=\!1,\!\ldots,\!M_b,\nonumber\\
        &\!\!\!{\rm P}_{u}|(\widehat{h}_{u,u}[i] + c_u[i])|^2\leq \lambda_u, \nonumber\\
        &\!\!\!\mathbb{E}\{\|\v_b[i] s_{b}[i]\|^2\}\leq {\rm P}_b,\,\,\text{and}\,\,\,
        \mathbb{E}\{|s_{u}[i]|^2\}\leq {\rm P}_u,\nonumber\\
        &\!\!\!\v_b[i] \in\mathbb{F}_{\rm TX}.
\end{align}
The first and second constraints impose the RX RF chain saturation thresholds $\lambda_{b}$ and $\lambda_u$ after analog cancellation at nodes $b$ and $u$, respectively. These saturation thresholds ensure successful reception of the training symbols and decoding of BS's data symbols. The next two constraints in \eqref{eq: optimization_eq} refer to the nodes' average transmit powers. The final constraint enforces the predefined analog codebook for the BS beamformer.
\begin{algorithm}[!t]
    \caption{Proposed FD massive MIMO SDDT Design}
    \label{alg:the_alg}
    \begin{algorithmic}[1]
        \renewcommand{\algorithmicrequire}{\textbf{Input:}}
       \renewcommand{\algorithmicensure}{\textbf{Output:}}
        \REQUIRE $\widehat{\H}_{b,b}[i]$, $\widehat{h}_{u,u}[i]$, ${\rm P}_b$, ${\rm P}_u$, $\widehat{\theta}_{\rm LoS}[i-1]$, and $M_b$.
        \ENSURE $\v_b[i],\c_b[i], c_u[i], \d_{b}[i]$, and $d_u[i]$.
        \STATE Obtain the DL channel estimate $\widehat{\h}_{u,b}[i]$ using $\widehat{\theta}_{\rm LoS}[i-1]$ as described in \eqref{eq: DL_est_chan}.
        \STATE Obtain the analog beamformer $\v_b[i] = \underset{\v \in\mathbb{F}_{\rm TX}}{\text{arg max}} \frac{|\widehat{\h}_{u,b}^{\rm T}[i]\v|^2}{\|\widehat{\H}_{b,b}[i]\v\|^2}$ using exhaustive search.
        \FOR{$n= 1,2,\ldots,M_b-1$}
            \STATE Set analog SI canceller $\c_b[i] = \begin{bmatrix}-[\widehat{\H}_{b,b}[i]\v_b[i]]_{1:n}\\
                        \mathbf{0}_{(M_b-n:M_b)}
                        \end{bmatrix}$.
            \STATE Set $c_u[i] = - \widehat{h}_{u,u}[i]$.           
            \IF {${\rm P}_b|[(\widehat{\H}_{b,b}[i]\v_b[i]\!+\!\c_{b}[i])]_{(j,:)}|^2\leq \lambda_b \, \forall j=1,\ldots,M_b$, and ${\rm P}_{u}|(\widehat{h}_{u,u}[i] + c_u[i])|^2\leq \lambda_u$}
            \STATE Output $\v_b[i]$, $\c_b[i]$, $c_u[i]$, $\d_{b}[i]=-(\widehat{\H}_{b,b}[i]\v_b[i] + \c_b[i])$, $d_u[i]=-(\widehat{h}_{u,u}[i]+ c_u[i])$, and terminate the iterations.
            \ENDIF
        \ENDFOR
        \STATE Set $\c_b[i] =-\widehat{\H}_{b,b}[i]\v_b[i]$ and $c_u[i] = - \widehat{h}_{u,u}[i]$.
        \IF {${\rm P}_b|[(\widehat{\H}_{b,b}[i]\v_b[i]\!+\!\c_{b}[i])]_{(j,:)}|^2\leq \lambda_b \,\forall j=1,2,\ldots,M_b$, and ${\rm P}_{u}|(\widehat{h}_{u,u}[i] + c_u[i])|^2\leq \lambda_u$}
            \STATE Output $\v_b[i]$, $\c_b[i]$, $c_u[i]$, $\d_{b}[i]=-(\widehat{\H}_{b,b}[i]\v_b[i] + \c_b[i])$, $d_u[i]=-(\widehat{h}_{u,u}[i]+ c_u[i])$, and stop the algorithm.
        \ELSE
            \STATE Output that the $\c_b[i]$ realizations or $c_u[i]$ do not meet the RX RF saturation constraints.
        \ENDIF
    \end{algorithmic}
\end{algorithm}

The optimization problem in \eqref{eq: optimization_eq} is a non-convex problem with coupling variables, hence, quite difficult to tackle. In this work, we solve it suboptimally using alternating optimization, leaving other possibilities for future work. First, we find the BS analog beamformer $\v_b[i]$ via the following problem:
\begin{equation}\label{eq: exhaust_search}
    \begin{split}
        \v_b[i] = \underset{\v \in\mathbb{F}_{\rm TX}}{\text{arg max}} \frac{|\widehat{\h}_{u,b}^{\rm T}[i]\v|^2}{\|\widehat{\H}_{b,b}[i]\v\|^2}.
    \end{split}
\end{equation}
The exhaustive search for this problem can be easily implemented by a simple look-up table given the estimated SI channel $\widehat{\H}_{b,b}[i]$ and the estimated DL channel $\widehat{\h}_{u,b}[i]$. Following the analog SI canceller structure in \cite{alexandropoulos2020full} and using $\v_b[i]$, we next seek the $N$-tap analog canceller $\c_b[i]$ with $1\leq N \leq M_b$ that satisfies the first threshold constraint. The single-tap canceller at node $u$ is obtained as $c_u[i] = - \widehat{h}_{u,u}[i]$. To maximize the signal-to-interference-plus-noise ratio, the digital cancellers $\d_b$ and $d_u$ are set as the respective complementary residual SI channels after analog SI cancellation. Our solution for the optimization problem \eqref{eq: optimization_eq} is summarized in Algorithm \ref{alg:the_alg}.

\begin{figure}[!tpb]
	\begin{center}
	\includegraphics[width=\linewidth]{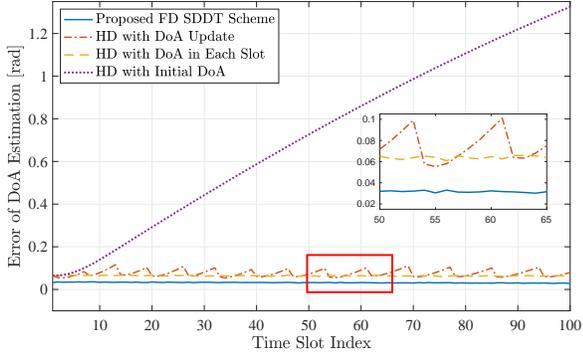}
	\caption{Error of LoS DoA estimation w.r.t the time slot index for the $64\times 2$ FD massive MIMO node $b$ communicating with the mobile single-antenna FD UE $u$ transmitting UL training symbols with $10$dBm transmit power.}
	\label{fig: DoA_error}
	\end{center}
	\squeezeupan
\end{figure}

\section{Numerical Results}
\squeezeupann
In this section, we present simulation results for the performance of the proposed SDDT scheme for FD mmWave massive systems in comparison with HD counterparts. 

\subsection{Simulation Parameters}\label{ssec: Sim_param}
We perform an extensive waveform simulation following the FD massive MIMO architecture illustrated in Fig.~\ref{fig: FD_HBF_Sys} when operating at mmWave frequencies, where a $64\times 2$ FD massive MIMO node $b$ is communicating with a single-antenna FD UE node $u$. We have assumed that the single TX RF chain at the BS node $b$ is connected to a ULA consisting of $N_b = 64$ antenna elements. In contrast, on the RX side, $M_b = 2$ RX antennas are connected to their dedicated RX RF chains. The UE node $u$ employs a non-ideal single-tap SI canceller, where the FD massive MIMO BS node $b$ deploys an $N=2$ taps analog SI canceller \cite{alexandropoulos2020full}. 
The multi-path UL and DL channels are simulated as mmWave channels at $28$ GHz each with one LoS and $4$ nLoS channel paths. Considering the distance $d_{\rm BS} = 100m$ of the BS node $b$ from the UE, the pathloss of the both UL and DL channels is assumed to be $100$dB with a $25$ dB Rician factor between the LoS and nLoS paths \cite{akdeniz2014millimeter}. In addition, the SI channels are modeled as Rician fading channels with a $\kappa$-factor of $35$dB and pathloss $40$dB \cite{alexandropoulos2017joint}. The RX noise floors at all nodes were assumed to be $-100$dBm. To this end, the RXs have an effective dynamic range of $60$dB provided by $14$-bit Analog-to-Digital Converters (ADC) for a Peak-to-Average-Power-Ratio (PAPR) of $10$ dB. Therefore, the residual SI power after analog SI cancellation at the input of each RX RF chain has to be below $-40$dBm to avoid signal saturation. 
For the considered FD massive MIMO architecture, we have assumed that the UE $u$ is moving at a constant velocity of $120$km/h for a duration of $100$ time slots, where each time slot is considered to be $T_s = 10$msec with $L=400$ symbols. For the BS analog beamformer, we have used a $6$-bit beam codebook based on the Discrete Fourier Transform (DFT) matrix.
We have used $1000$ independent Monte Carlo simulation runs to calculate the performance of all considered DoA estimation and data transmission designs.

\begin{figure}[!tpb]
	\begin{center}
	\includegraphics[width=\linewidth]{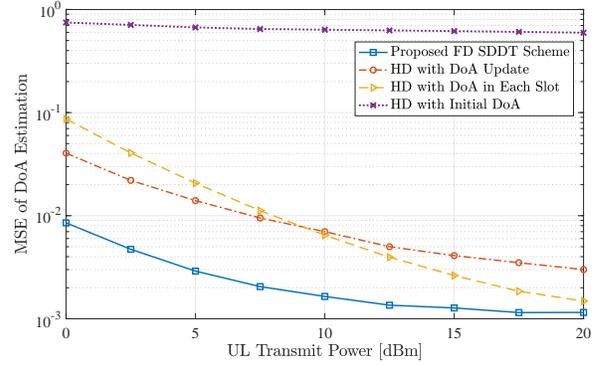}
	\caption{MSE of the DoA estimation with respect to UL transmit power in dBm for the $64\times 2$ FD massive MIMO node $b$ communicating with the single-antenna FD UE $u$ transmitting training symbols in the UL direction.}
	\label{fig: MSE_DoA}
	\end{center}
	\squeezeupan
\end{figure}
\vspace{-0.1cm}
\subsection{Compared HD Massive MIMO Designs}\vspace{-0.1cm}
We compare the performance of the proposed FD-based SDDT scheme with three HD DoA-assisted DL analog beam management techniques. First, we consider the ``HD with Initial DoA" case, where DoA is estimated using the UL training symbols only in the first time slot, and for the rest of the $99$ time slots, the same DoA is utilized for finding the DL analog beam. Secondly, we simulate the ``HD DoA in Each Slot" scheme, where, for every time slot, a fraction of the UL training symbols is utilized for DoA estimation, and the rest of the symbols are dedicated for data transmission with analog BF. Finally, we consider the ``HD DoA with Update" technique, where DL analog BF in each time slot is performed using the DoA estimation of the previous time slot, unless more than $3$dB SNR loss occurs in the received DL signal, which triggers the initialization of a DoA-based analog beamformer update. In addition to the latter HD cases, we have simulated the achievable DL rate for the proposed FD-based SDDT scheme with ideal DoA estimation.

\vspace{-0.1cm}
\subsection{DoA Estimation Error and Average DL Rate}\vspace{-0.1cm}
In Fig.~\ref{fig: DoA_error}, the LoS DoA estimation error in radians is illustrated with respect to the time slot index. We have considered that the FD massive MIMO node $b$ communicates with the mobile single-antenna FD UE $u$, while the latter sends training symbols in the UL for DoA estimation with power $10$dBm. It is shown in the figure that the ``HD with Initial DoA" scheme provides higher DoA estimation error with increasing time slot index. This happens because it only estimates the DoA in the first time slot. As the UE moves, the initial DoA becomes outdated, and hence, a DoA estimation as well as an analog beam update are required. As illustrated in the inset plot, the ``HD DoA with Update" scheme performs an LoS DoA update after reaching approximately $0.1$rad estimation error, while the ``HD DoA in Each Slot" results in a steady error as the estimation is performed in each time slot. It is, however, evident from Fig.~\ref{fig: DoA_error} that the proposed FD-based SDDT scheme with a $2$-tap analog canceller provides substantially lower DoA estimation error compared to the HD cases across all time slots for the considered $10$dBm UL transmit power. In Fig.~\ref{fig: MSE_DoA}, we plot the Mean Squared Error (MSE) of the DoA estimation for all considered schemes as a function of the UL transmit power in the range $0-20$dBm. It is clear from the figure that the MSE of the DoA estimation reduces with increasing UL transmit power. This happens because the received SNR at the BS increases. It is also shown that the proposed scheme outperforms all  considered HD-based counterparts.

\begin{figure}[!tpb]
	\begin{center}
	\includegraphics[width=\linewidth]{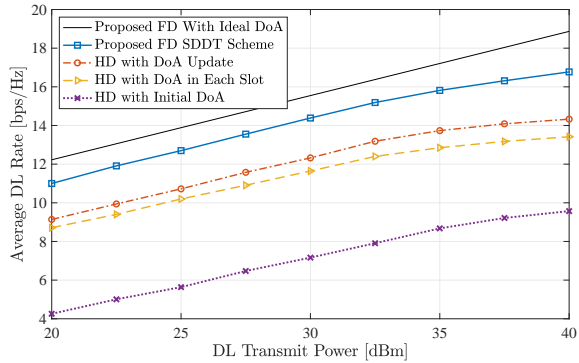}
	\caption{Achievable DL rate with respect to DL transmit power in dBm for the $64\times 2$ FD massive MIMO node $b$ communicating with the mobile single-antenna FD UE $u$ sending training symbols with $10$dBm transmit power.}
	\label{fig: DL_rate}
	\end{center}
	\squeezeupan
\end{figure}

Figure~\ref{fig: DL_rate} illustrates the achievable DL rate of the proposed SDDT scheme as a function of the DL transmit power, considering that the $64\times 2$ BS node $b$ is transmitting data in the DL via analog BF to the constant-velocity mobile UE node $u$, while receiving UL training symbols with $10$dBm transmit power. It is demonstrated that the ``HD with Initial DoA" provides the lowest DL rate for all DL transmit power levels, as it suffers from the highest DoA estimation error. It is also shown that, although the ``HD DoA with Update" scheme achieves higher DoA estimation error compared to the ``HD DoA in Each Slot" approach, as depicted in Fig.~\ref{fig: DoA_error}, it achieves higher DL rate across all DL transmit powers. This is due to the fact that the later HD DoA-assisted beam management scheme utilizes $10\%$ symbols in each time slot for DoA estimation, whereas the ``HD DoA with Update" approach is capable of sending data symbols for the whole time slot when it doesn't have to update the DoA estimation. It is finally evident that the proposed scheme with $2$-tap analog cancellation achieves a higher DL rate compared to all the HD-based schemes fo all DL transmit powers. In particular, for the large DL transmit power of $40$dB, the proposed scheme for FD mmWave massive MIMO systems results in $120\%$ of the achievable DL rate with the best HD-based counterpart.
\section{Conclusion}
In this paper, we presented a novel DoA-assisted analog beam management scheme for FD mmWave massive MIMO systems. We considered an FD massive MIMO BS with an analog beamformer serving a mobile single-antenna FD user moving at a constant velocity. By adopting the MUSIC DoA estimation technique as an example, we presented a joint design of the DoA-assisted analog beamformer and A/D SI cancellation at the BS node maximizing the DL rate. Our performance evaluation results considering a realistic mmWave channel model demonstrated the superior achievable rates of the proposed FD-based SDDT scheme. In future work, the simultaneous DoA estimation and data transmission protocol will be considered for multi-user FD massive MIMO systems with both A/D precoders and combiners at all involved nodes.
\section*{Acknowledgments}
This work was partially funded by the National Science Foundation CAREER award \#1620902.


\bibliographystyle{IEEEtran}
\bibliography{IEEEabrv,ms}

\end{document}